\documentclass[conference]{IEEEtran}

\usepackage{amssymb}
\usepackage{graphicx}
\usepackage{epstopdf}
\usepackage{gensymb}
\usepackage{color}
\usepackage{amsmath}
\usepackage{bm}
\usepackage{etoolbox}
\usepackage{cite}
\usepackage{array}
\usepackage{booktabs}
\usepackage{multirow}
\usepackage{comment}
\usepackage{url}

\usepackage{threeparttable} 
\usepackage{mathtools}
\usepackage{pifont}

\IEEEoverridecommandlockouts
\IEEEpubid{\makebox[\columnwidth]{\copyright~2019 IEEE. Permission from IEEE must be obtained except personal usage.} \hspace{\columnsep}\makebox[\columnwidth]{ }}

\newcommand{\damith}[1]{\textcolor{blue}{Damith: #1}}

\begin{document}

\title{Hash Functions and Benchmarks for Resource Constrained Passive Devices: A Preliminary Study}

\author{
   \IEEEauthorblockN{Yang Su\IEEEauthorrefmark{1}, Yansong Gao\IEEEauthorrefmark{2}\IEEEauthorrefmark{3},  Omid Kavehei\IEEEauthorrefmark{4}, and Damith C.~Ranasinghe\IEEEauthorrefmark{1}}
  \IEEEauthorblockA{\textit{\IEEEauthorrefmark{1}Auto-ID lab, School of Computer Science, The University of Adelaide}, Australia}
   \IEEEauthorblockA{\textit{\IEEEauthorrefmark{2} Data61, CSIRO}, Sydney, Australia}
   \IEEEauthorblockA{\textit{\IEEEauthorrefmark{3} School of Computer Science and Engineering, Nanjing University of Science and Technology}, China}
   \IEEEauthorblockA{\textit{\IEEEauthorrefmark{4} School of Electrical and Information Engineering, The University of Sydney}, Sydney, Australia
  \\ yang.su01@adelaide.edu.au; yansong.gao@njust.edu.cn;  \\omid.kavehei@sydney.edu.au; damith.ranasinghe@adelaide.edu.au}
}


%



\maketitle
\begin{abstract}
Recently, we have witnessed the emergence of intermittently powered computational devices, an early example is the Intel WISP (Wireless Identification and Sensing Platform). How we engineer basic security services to realize mutual authentication, confidentiality and preserve privacy of information collected, stored and transmitted by, and establish the veracity of measurements taken from, such devices remain an open challenge; especially for batteryless and intermittently powered devices.  
While the cryptographic community has significantly progressed lightweight (in terms of area overhead) security primitives for low cost and power efficient hardware implementations, lightweight software implementations of security primitives for resource constrained devices are less investigated. Especially, the problem of providing security for \textit{intermittently powered} computational devices is unexplored. In this paper, we illustrate the unique challenges posed by an emerging class of intermittently powered and energy constrained computational IoT devices for engineering security solutions. We focus on the construction and evaluation of 
a basic hash primitive---both existing cryptographic hash functions and non-cryptographic hash functions built upon lightweight block ciphers. We provide software implementation benchmarks for \textit{eight} primitives on a low power and resource limited computational device, and outline an execution model for these primitives under intermittent powering. 
\end{abstract}


\begin{IEEEkeywords}
Hash functions, Benchmarks, Ultra low power microcontrollers, Power harvesting, Computational RFID, Intermittently powered devices, Energy constrained devices, Intermittent execution model
\end{IEEEkeywords}

\IEEEpeerreviewmaketitle

\section{Introduction}
The realization of new applications from studying the behaviors of bees~\cite{henry2012common} to healthy aging of our species~\cite{wickramasinghe2017sequence} with tiny computing and sensing devices are providing the impetus for new growth areas in the field of Internet of Things (IoT). However, provisioning of fundamental security services such as authentication, confidentiality and message integrity remains a challenge for resource limited environments such as with IoT devices. 

\begin{figure}[t]
    \centering
    \includegraphics[width=.8\linewidth]{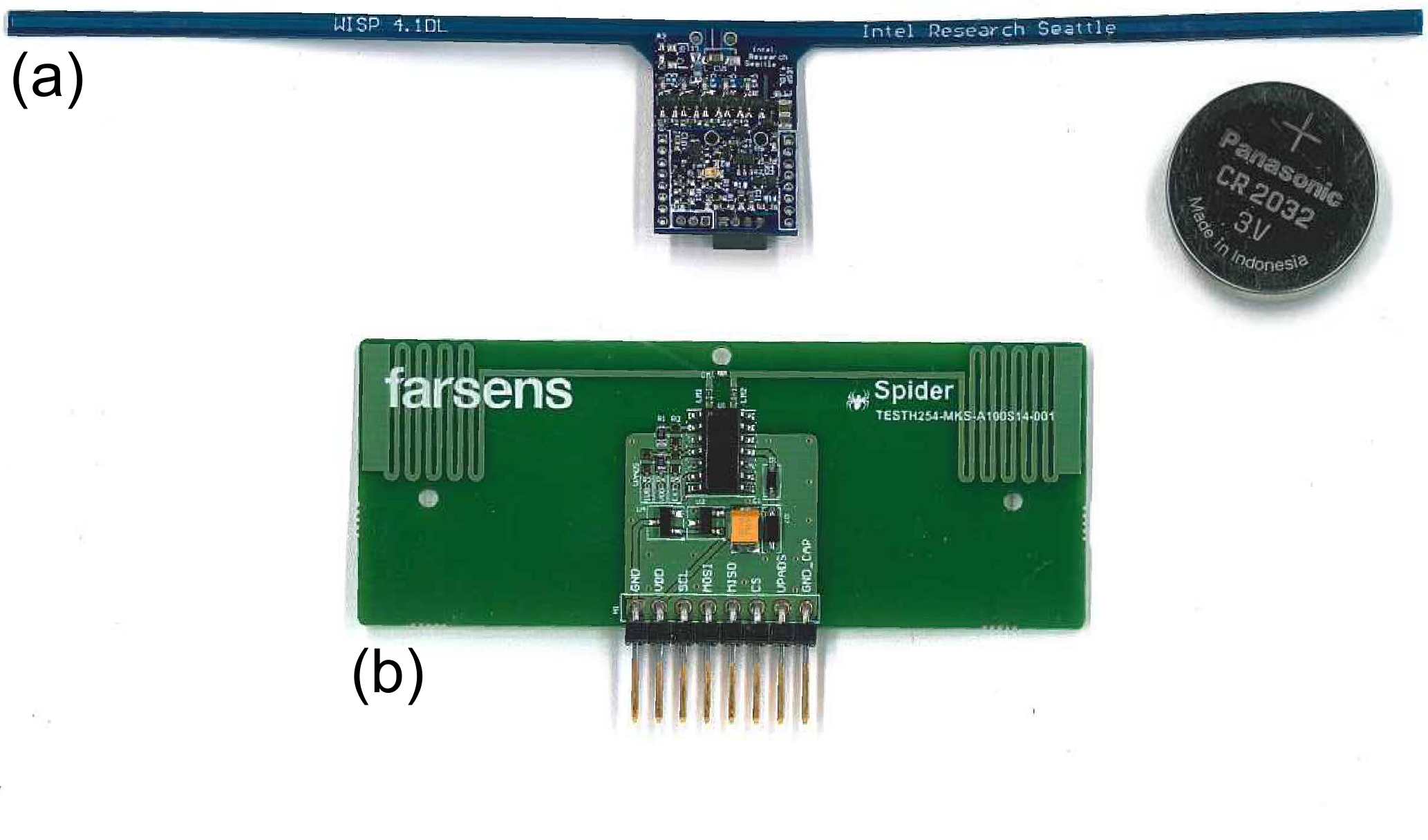}
    \caption{Examples of intermittently powered computations and energy constrained devices operating on harvested power. (a) Intel CRFID device: WISP4.1DL\cite{sample2008design,aaronparks2017}; and (b) Farsens Spider CRFID device\cite{farsens2015}.} 
    \label{fig:WISPs}
\end{figure}

Computational batteryless or passive technologies, exemplified the Intel WISP and the Farsens Spider illustrated in Fig.~\ref{fig:WISPs}, are a class of emerging low power embedded systems and communication technologies attracting increasing attention from both academia and industry sectors driven by their ability to operate, potentially, indefinitely and often in deeply embedded applications~\cite{liu2017novel} where battery replacements are not practicable. The challenge of providing basic security services to this emerging class of devices employing ultra low power computing platforms is significantly more difficult as a result of intermittent powering, and severe computation and energy limitations. 

To date, mostly hardware cost minimization---implementing security primitives with, for example, less transistors for application specific integrated circuits (ASICs)---is considered for realization of lightweight security primitives. However, the problem of supporting security services for tiny computational platforms exemplified by low power and reduced instruction set microcontrollers increasingly being used to realize Internet of Things (IoT) devices has received less attention; while, the problem of providing security for \textit{intermittently powered} computational devices is unexplored. 

In this paper, we investigate the unique challenges posed by an emerging class of intermittently powered computations and energy constrained computational devices for engineering security solutions. While encryption and decryption primitives for ultra low power microcontrollers have received some attention~\cite{cazorla2015survey,cazorla2013survey}, we focus on the unexplored software implementation and evaluation of 
hash primitives---both existing hash functions and non-cryptographic hash functions built upon lightweight block ciphers---on ultra low power microcontrollers suitable for energy constrained  devices. \textit{For the first time} and to the best of our knowledge, we provide benchmarks of hash functions expected to yield a lightweight implementation in a reduced instruction set architecture (RISC) microcontroller. 
It is hoped that such benchmarks will provide a useful guide to  engineering crytographic solutions for resource constrained devices. We also briefly discuss an execution model for hash primitives based on~\cite{su2018secucode} for operating under intermittent powering. We begin with an illustration of the unique challenges posed by intermittently powered and resource limited devices.




\section{Intermittently Powered Pervasive Devices}

In contrast to depleting chemical energy in a battery, energy harvesting and battery free devices scavenge power from ambient sources, such as radio waves, vibrations, and solar radiation or are remotely energized through wireless powering.  A generic architecture of a device operating on harvested energy is shown in Fig.~\ref{fig:HarvestArchi-curve} (a). An input oscillating voltage is rectified---and often boosted through a charge pump---to $V_{\rm{cap}}$ and regulated to voltage $V_{\rm{reg}}$ used to operate a load, e.g., an ultra low power microcontroller.

\subsection{Intermittent Powering}
\begin{figure}[t]
    \centering
    \includegraphics[width=1\linewidth]{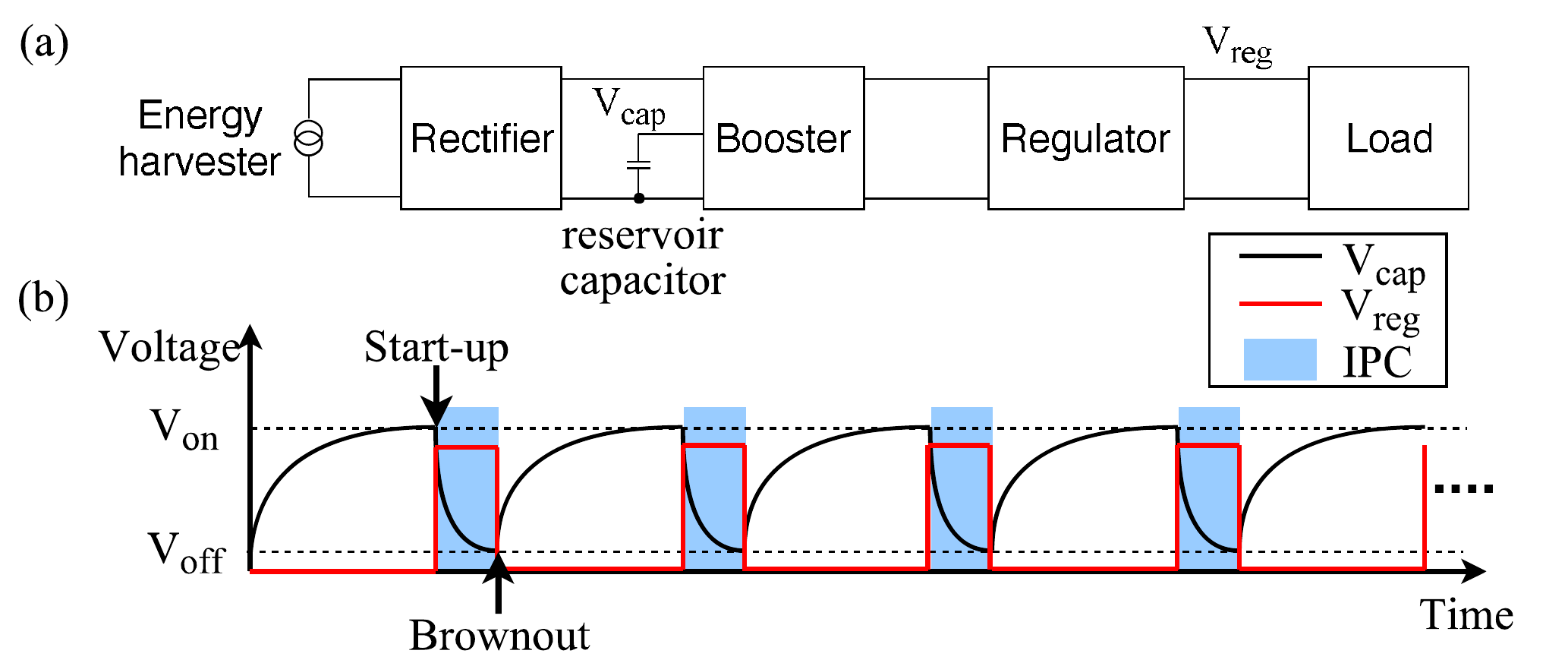}
    \caption{(a) The generic architecture of an energy harvesting device and (b) the intermittent power cycle (IPC) resulting from the reservoir capacitor's charge-discharge characteristics.}
    \label{fig:HarvestArchi-curve}
\end{figure}

A typical battery free device operation can be described by the intermittent power cycle (IPC) illustrated in Fig. \ref{fig:HarvestArchi-curve}(b). The voltage $V_{\rm{cap}}$ across the reservoir capacitor gradually ramps up until $V_{\rm{on}}$, the startup voltage of the booster circuit. As the regulator delivers the necessary power to the load, the charge on the reservoir capacitor and hence $V_{\rm{cap}}$ is rapidly drawn down, subsequently, the booster cuts off when $V_{\rm{cap}}$ drops below $V_{\rm{off}}$; that is when the rectifier circuit cannot replenish the reservoir capacitor faster than its draw down. When $V_{\rm{cap}}$ falls below $V_{\rm{off}}$, a \textit{brownout} event occurs resulting in complete state loss and restarting of the microcontroller---Load in Fig.~\ref{fig:HarvestArchi-curve}. Hence, only within the period highlighted in light blue, can the Load be actually powered. Such a period is called the intermittent power cycle (IPC), which is short and fragmented over time in practice~\cite{su2018secucode}.

\begin{figure}[t]
    \centering
    \includegraphics[width=1\linewidth]{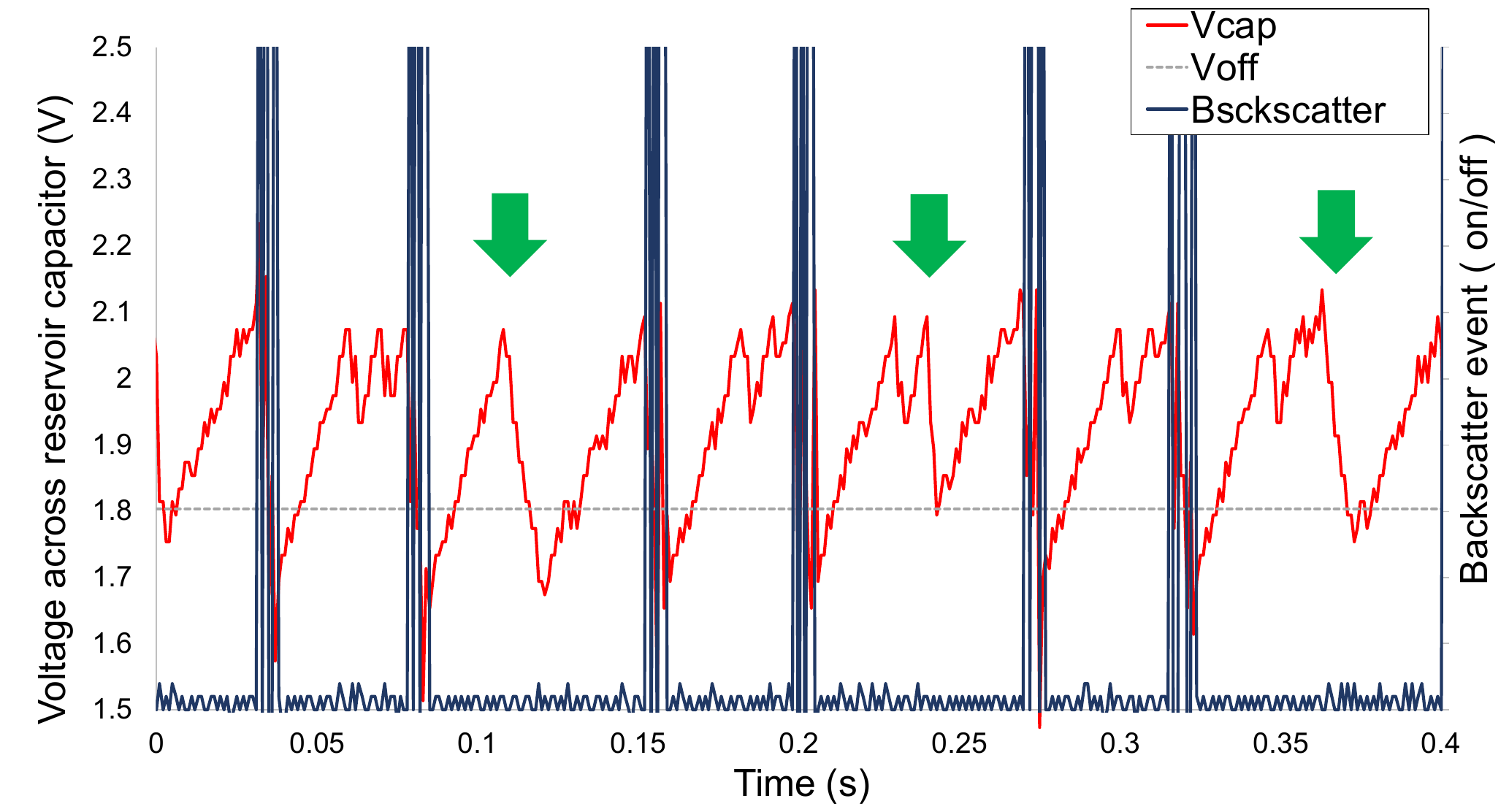}
    \caption{Sudden power loss of a CRFID device with two sensors. During the sensor sampling operation, multiple occurrences of power loss (brownout events) are observed (indicated by the green arrows).}
    \label{fig:Sudden-Power-loss}
\end{figure}

Fig.~\ref{fig:Sudden-Power-loss} demonstrates an example of intermittent powering affects on the operation of a batteryless device running on harvested power. 
In this example from~\cite{su2015investigating}, a WISP4.1DL CRFID device is wirelessly powered and executing firmware to sample two on-board sensors---an accelerometer and a barometer---and backscattering data to an RFID reader to support sensor data retrieval; the sudden power loss event during the sensor sampling process is highlighted by green arrows. All operations must be completed prior to a brownout event or within an IPC; otherwise the device will loose its state immediately. This problem becomes more severe in the engineering of a security layer given the computational complexity and power consumption of security primitives.

Given a power loss event, the device will be rebooted once the reservoir capacitor is charged and the control flow restarts from the entry point of the application instead of the moment before power failure. 
In the worst case, the application program always encounters power failures, and may never completely execute a task.

\subsection{Available Clock Cycles}
Given intermittent powering, devices are also limited by the computations---measured by clock cycles---that can be executed before a brownout event. Theoretically, available energy to the load $\mathcal{E}_{\rm laod} $ within one single IPC can be expressed by \eqref{eqn:availablePower}, with $\mathcal{E}_{\rm RF}$ the energy available from the energy harvester during an IPC, and $C$ the reservoir capacitor's  capacitance. This capacitor starts discharging when its terminal voltage reaches $V_{\rm on}$, and stops discharging below $V_{\rm off}$. Since $C$, $V_{\rm on}$ and $V_{\rm off}$ are all constant, when $\mathcal{E}_{\rm RF} \ll \frac{1}{2}\cdot C \cdot (V_{\rm on} - V_{\rm off})^2$, the energy available to the load is dominated by the size of the reservoir capacitor. 
Otherwise, the reservoir capacitor is replenished before getting discharged and  the device could operate continuously.

\begin{equation}
\label{eqn:availablePower}
	\mathcal{E}_{\rm load}=\frac{1}{2}\cdot C \cdot (V_{\rm on}- V_{\rm off})^2
\end{equation}

Total available clock cycles $ N_{\rm ck}$ that the CPU could execute before power loss can be expressed by:
\begin{equation}
\label{eqn:availableCLK}
	N_{\rm ck} = \frac{\mathcal{E}_{\rm load}}{P_{\rm load}} \cdot f_{\rm CPU}
\end{equation}
with $P_{load}$ the average power of the load including the computational elements and sensors, and $f_{CPU}$ the selected operating frequency of the CPU.

\begin{figure}[t]
    \centering
    \includegraphics[trim=0cm 0cm 0cm 1.4cm,clip,width=0.8\linewidth]{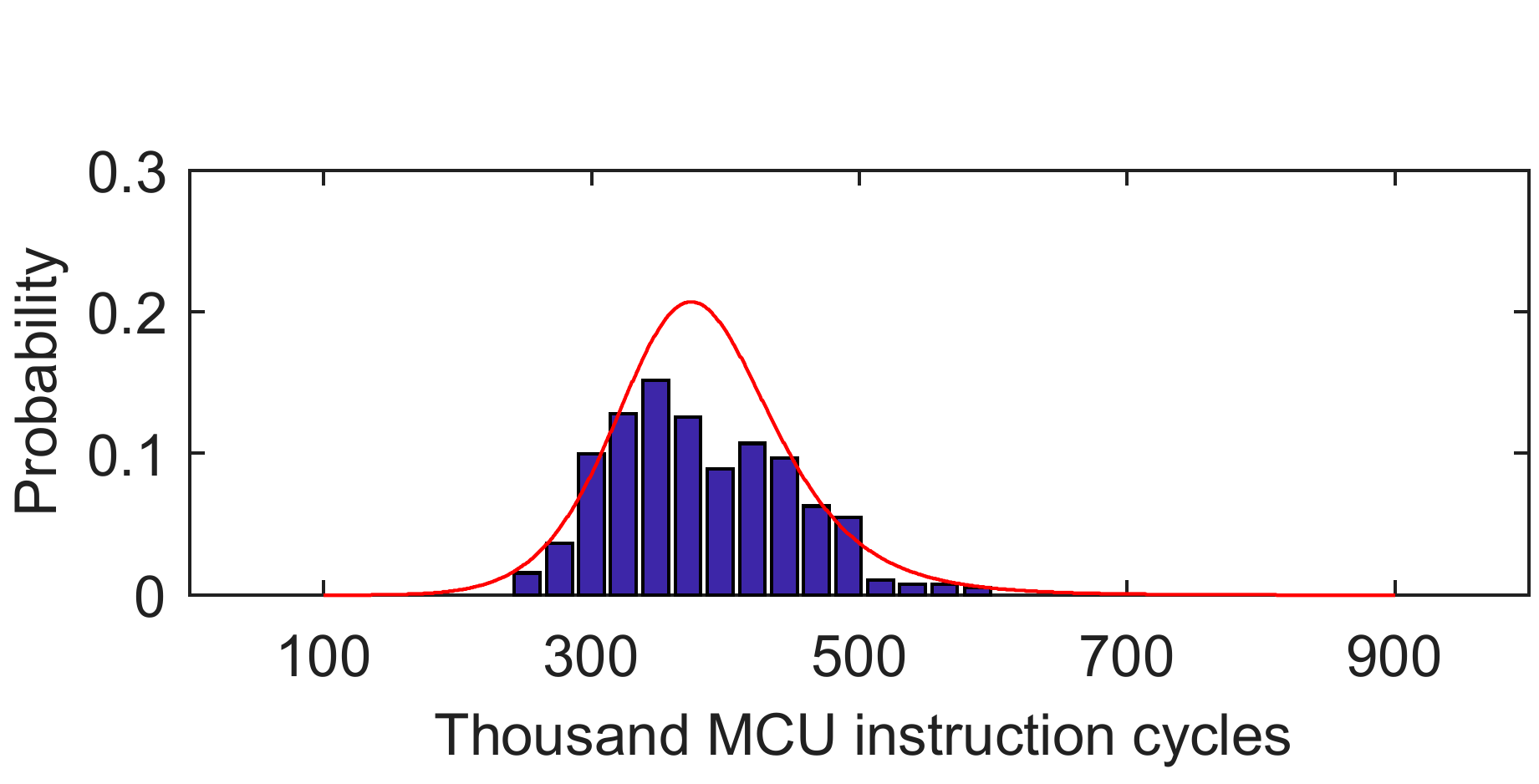}
    \caption{An experimental evaluation of the available clock cycle distribution within a IPC of a batteryless WISP. Data was collected at 50~cm from a radiating RFID reader antenna.}
    \label{fig:Available_CLK}
    \vspace{-10pt}
\end{figure}

To quantify the available clock cycles within one IPC, we follow the method and setup described in~\cite{tan2016wisent}. In our measurement, a hash function is executed \textit{without} any RFID communications to reduce the influence from uncontrolled variables---such as energy loss during backscattering, clock frequency change when executing RFID protocol commands and random inactive times due to the random time slot selected for communicating with an RFID reader---on the measurements. We can see the typical distribution of clock cycles under intermittent operating conditions in Fig.~\ref{fig:Available_CLK}. We can see that due to reasons such as variable energy available from the energy harvester $\mathcal{E}_{\rm RF}$ and the ability of power harvesting circuits to replenish the reservoir capacitor, the available number of clock cycles vary in practice.

\section{Related Work}
Research studies have benchmarked block ciphers~\cite{cazorla2015survey} with clearly defined testing frameworks and publicly available source code aimed at embedded systems including AVR8051~\cite{rinne2007performance} and ATinny45\cite{hatzivasilis2018review}. 
However, we observe that the method of logging clock cycles is not explicitly mentioned; for example, we can see differences in reported results between \cite{cazorla2013survey} and \cite{cazorla2015survey}.  

Although, benchmarks such as the recent work in~\cite{hatzivasilis2015password} which consists of 22 password hashing functions, are reported for desktop platforms, to our knowledge, there are no software implementation benchmarks of hash functions on resource limited microcontrollers. 
For resource constrained devices, hash function evaluations have focused on hardware implementations such as on ASIC (application specific integrated circuits) such as RFID ICs (integrated circuits) or FPGA (field-programmable gate arrays) platforms as in~\cite{bogdanov2008hash,feldhofer2006case}. 
In our study, we focus on recent ultra low power computing platforms such as the the MSP430 series from Texas Instruments where the need is to build lightweight---both in terms of computations and energy---security primitives such as hash functions in software as opposed to hardware. 

\begin{table}[]
\centering
\caption{A List of Hash Functions Evaluated}
\label{Tab:TestHash}
\resizebox{\linewidth}{!}{
\begin{tabular}{llllll}
\hline
Name       & Structure & digest size (bits) & Attack complexity & Attack ref \\\hline
BLAKE2s~\cite{aumasson2013blake2} & HAIFA  & 256            &C=$2^{184}$ &\cite{hao2014boomerang}\\
MD5~\cite{rivest1992md5}&M-D& 128  & C=$2^{18}$ (2-block),P=$2^{123.4}$&\cite{xie2013fast},\cite{sasaki2009finding}\\
SHA-1~\cite{eastlake2001us} &M-D&160          &C=$2^{63.1}$                   &\cite{merrill2017better}\\
SHA-3~\cite{dworkin2015sha} & Sponge & 256 & C=$2^{85.3}$,P=$2^{128}$ & \cite{unruh2016collapse}~\cite{unruh2017collapsing}\\
DM-PRESENT~\cite{poschmann2009lightweight} &DM&64           &C=$2^{29.18}$ &\cite{koyama2012multi}\\
DM-SPECK           &M-D&128             &                 &                  \\
MMO-SPECK           &M-D&128             &                 &                  \\
MP-SPECK           &M-D&128             &                 &                  \\
\hline
\end{tabular}}
\begin{tablenotes}
	\item{Best known attacks against the cryptographic hash functions at the time of writing. M-D: Merkle-Damg{\aa}rd construction, C: Collision resistance, P:preimage resistance}
\end{tablenotes}
\end{table}

\section{Evaluated Hash Functions}\label{Sec:fullHashList}
 We refer to existing hash functions as cryptographic hash functions since their security has generally been well assessed. In general, one-way compression functions built from block ciphers can also be used to construct hash functions---for a brief introduction see~\cite{bartkewitz2009building}. Among the methods used for building one-way compression functions are the well known constructions of Davies-Meyer (DM), Matyas-Meyer-Oseas (MMO), and Miyaguchi-Preneel (MP)\footnote{Although there are other methods as in~\cite{bogdanov2008hash}, in this preliminary study, we are interested in comparing the representative implementation overhead from constructed hash functions from lightweight block ciphers with cryptographic hash functions on a resource limited device. Therefore we will not discuss  details of hash function designs and limit ourselves to DM, MMO and MP.}.  In this context, the computational complexity of the constructed hash function increases with the complexity of the underlying block cipher. Therefore, we selected the \textbf{SPECK} block cipher since it showed the least overhead---see evaluations in Table~\ref{tab:cipherSize}. We refer to hash functions we have derived based on compression functions built from lightweight block ciphers using the popular Merkle-Damg{\aa}rd construction method as non-cryptographic hash functions. The full list of all hash functions evaluated in this work is summarized in Table~\ref{Tab:TestHash}. Here, BLAKE2s, MD5, and DM-PRESENT 
 are cryptographic hash functions. In the following, we describe the construction of non-cryptographic hash functions: DM-SPECK, MMO-SPECK, and MP-SPECK. 
\vspace{1mm}



\begin{figure}[t]
    \centering
    \includegraphics[width=\linewidth]{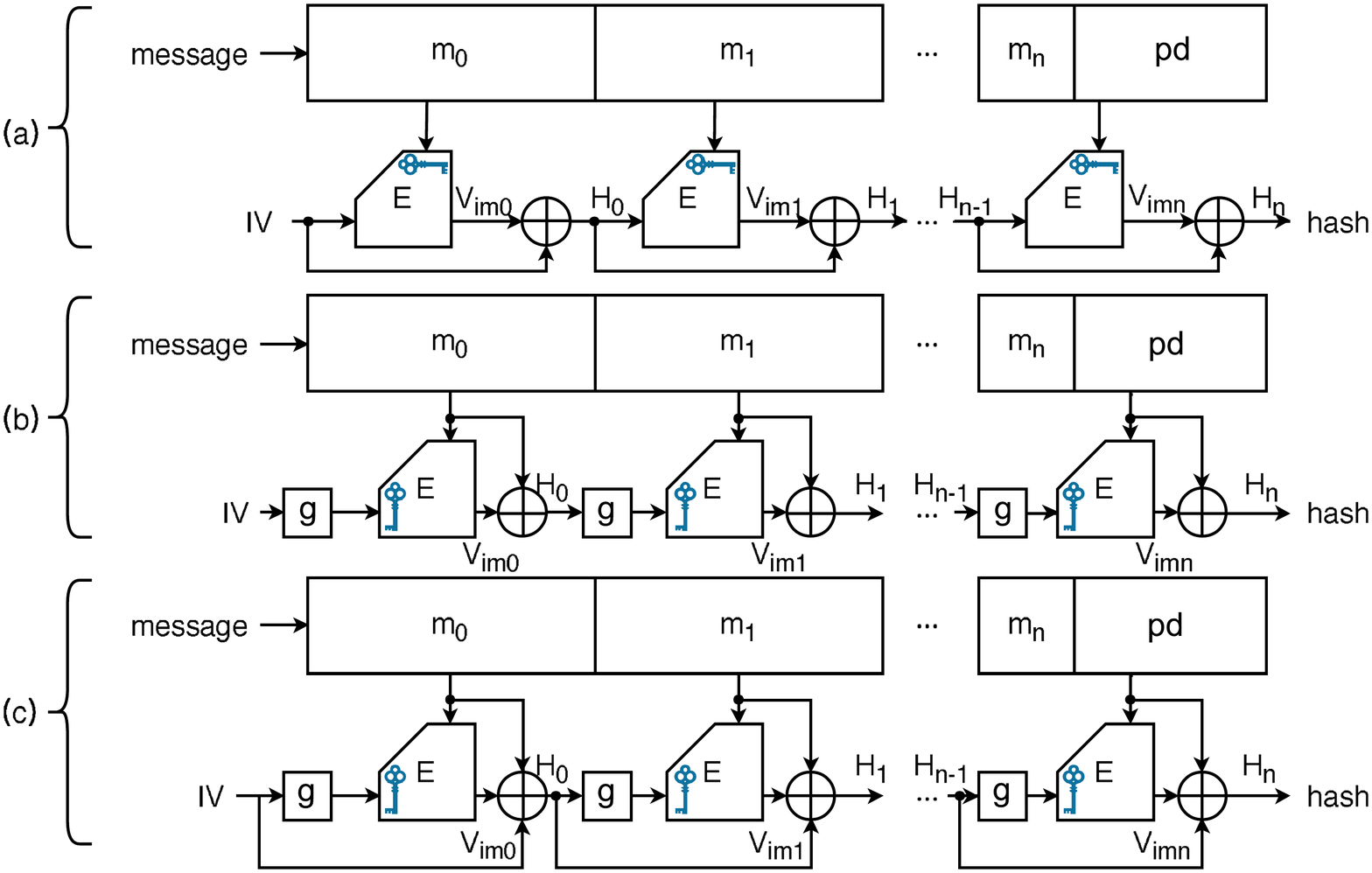}
    \caption{construction of hash functions using (a) Davies-Meyer (DM), (b) Matyas-Meyer-Oseas (MMO) and (c) Miyaguchi-Preneel (MP) configuration.}
    \label{fig:MD_construction}
\end{figure}

\noindent\textbf{DM-SPECK: }The configuration of Davies-Meyer (DM) compression function based construction is depicted in Fig.~\ref{fig:MD_construction} (a). The input message $m$ is first sliced into $n$ blocks with $k$-bits each---$k$ is the key size of the cipher $E$. In case the message is not an integer multiple of $k$, zero-padding  is appended to the end of the message $m$. At each DM compression stage, the output hash value of the previous stage $H_{i-1}$ is fed into $E$ to obtain the intermediate value $V_{im}$. Subsequently, $V_{im}$ is XORed with $H_{i-1}$ to obtain $H_i$. A fixed initialization vector (IV) is input at the initial stage.
\vspace{1mm}

\noindent\textbf{MMO-SPECK: }The Matyas-Meyer-Oseas (MMO) based hash function configuration is depicted in Fig. \ref{fig:MD_construction} (b). Unlike the DM construction, the message fragment $m_i$ is fed to the plaintext port of the cipher $E$ instead of the key port. The output from the previous stage is first re-arranged through the $g$ function to match the size of key port of cipher $E$. The output of $E$ is XORed with message $m_i$ to produce the next value $H_{i}$.
\vspace{1mm}

\noindent\textbf{MP-SPECK: }The Miyaguchi-Preneel (MP) is a variant of the MMO configuration, where the next hash value $H_i$ is obtained by XORing the previous hash value $H_{i-1}$ with $V_{im}$ and $m_i$.
\vspace{1mm}

Notably, when dealing with necessary $a$-size-to-$b$-size bit length conversions in the implementation of the $g$ function, we can follow two types of implementations: i)~padding zeros; and ii)~duplicating inputs to perform a $n$-to-$2n$ mapping.


\section{Benchmarking}

\subsection{Hash Function Implementations} \label{hardware-platform}
We selected the MSP430FR5969 microcontroller (MCU) for evaluating the performance of software based hash function implementations. We selected this 16-bit ultra low power MCU due to the following attributes: 
\begin{itemize}
\item 100~$\mu$A/MHz active mode current: low energy consumption per clock cycle and, thus, the possibility to execute more instructions within one IPC.
\item 0.25 $\mu$A sleep mode current with an RTC (real-time clock): low power deep sleep mode can be used to accumulate harvested energy for future computationally intensive tasks whilst preserving state.
\item 64~KB of on-board FRAM (ferroelectric random access memory) based non-volatile memory: FRAM can be operated under a low supply voltage compared to more popular non-volatile memories such as EEPROM (electrically erasable programmable read-only memory). Thus, more suitable for energy limited platforms.
\item Good compiler tool chain support.
\end{itemize}

All ciphers and hash implementations in this paper are coded under the C99 standard. Majority of the code used is based on an open source repository from~\cite{kevinmarquet2014} after fixing certain software bugs. 
The test environment is the TI CCS (Code Computer Studio) 7.2.0 with candidate code being downloaded to a MSP430FR5969 LaunchPad Evaluation Kit via a USB interface. TI CCS provides a built-in GCC tool chain for the hardware kit, which includes the GNU 6.4.0.32\_win32 compiler.


\subsection{Benchmark Metrics}
The clock cycles required to complete a hash function operation is the most significant measure. 
The clock cycles are read out with the Profile Clock tool supported in the CCS environment. For  a fair comparison, the clock cycles are normalized as clock cycles per byte. Two different input message sizes are considered: a short message~(\textit{Short}) of 10 Bytes (notably, MD5 has 64 byte block size, and hash functions constructed from SPECK has a 16 byte block size), and a long message~(\textit{Long}) of 1,280 Bytes. The short message can be digested within one single hash iteration. The dominant factors in \textit{Short} message performance is the initialization string padding and data input/output. In the \textit{Long} message test, the message size is much larger than the block size and the dominant factor influencing performance is the  multiple compression processes. 
 
Memory footprint comprising of ROM usage and RAM usage is the other performance metric. ROM usage is indicative of the code size, and the RAM usage represents the size of the internal state necessary for the algorithm. The code size was read from the \texttt{.text} block in FRAM using the Memory Allocation tool in CCS, while the internal state was manually counted for any local variables declared within the algorithm routine.



 
\subsection{Block Cipher Performance}


The performance results of tested block ciphers are detailed in Table~\ref{tab:cipherSize}. Overall, we can observe that SPECK demonstrates the best performance. This is the main reason that it is chosen for constructing the non-cryptographic hash functions.

 \begin{table}[h]
\centering
\caption{Block Cipher Performance with Compiler Setting (-Os)}
\label{tab:cipherSize}
\resizebox{\columnwidth}{!}{%
\begin{tabular}{llllll}
\hline
Name     & Block size & Cycle count & Cycle per byte & ROM usage & RAM usage \\\hline
AES128   & 128        & 26822       & 1676           & 1136      & 18        \\
Camellia & 128        & 42959       & 2685           & 19866     & 268       \\
CLEFIA   & 128        & 70658       & 4416           & 1784      & 292       \\
LBlock   & 64         & 18769       & 2346           & 704       & 10        \\
LEA      & 128        & 16646       & 1040           & 1678      & 44        \\
PRINCE   & 64         & 14916       & 1865           & 1006      & 22        \\
SEA      & 96         & 65177       & 5431           & 660       & 18        \\
SIMON    & 64         & 13198       & 9939           & 326       & 5         \\
\textbf{SPECK}    & \textbf{64}         & \textbf{9939}        & \textbf{1242}           & \textbf{306}       & \textbf{1}         \\
XTEA     & 64         & 24423       & 3053           & 410       & 24       \\\hline
\end{tabular}
}
\end{table}
 
 \begin{table*}[]
\caption{Evaluation Results of Hash Functions}
\label{tab:hash_result}
\resizebox{\textwidth}{!}{
    \begin{tabular}{|l|l|l|l|l|l|l|l|l|l|l|l|l|l|l|l|}
    \hline
    \multirow{3}{*}{Hash} & \multicolumn{15}{c|}{Security Performance$^1$}  \\ \cline{2-16} 
                          & \multirow{2}{*}{Avalanche} & \multicolumn{2}{c|}{Cyclic} & \multicolumn{2}{c|}{TwoBytes}                          & \multicolumn{2}{c|}{Differential} & \multicolumn{2}{c|}{Sparse}                            & \multicolumn{2}{c|}{Permutation}& \multicolumn{2}{c|}{Window}& \multicolumn{2}{c|}{Zeros}      \\ \cline{3-16}
                          
                          & & \multicolumn{1}{c|}{Col.} & \multicolumn{1}{c|}{Dist.} & \multicolumn{1}{c|}{Col.} & \multicolumn{1}{c|}{Dist.} &  \multicolumn{1}{c|}{Col.} & \multicolumn{1}{c|}{Dist.} & \multicolumn{1}{c|}{Col.} & \multicolumn{1}{c|}{Dist.} & \multicolumn{1}{c|}{Col.} & \multicolumn{1}{c|}{Dist.} & \multicolumn{1}{c|}{Col.} & \multicolumn{1}{c|}{Dist.} & \multicolumn{1}{c|}{Col.} & \multicolumn{1}{c|}{Dist.} \\ \hline
    MD5       & 0.78\% & 0 & 0.049\% & 0 & 0.493\% & 0 & N/A & 0 & 0.127\% & 0 & 0.096\% & 0 & N/A & 0 & 0.493\% \\ \hline
    DM-SPECK  & 0.84\% & 0 & 0.037\% & 66369615 & 36.633\% & 0 & N/A & 0 & 0.103\% & 2392648 & 1.548\% & 0 & N/A & 61440 & 66.149\% \\ \hline
    MMO-SPECK & 0.81\% & 0 & 0.049\% & 66369615 & 36.621\% & 0 & N/A & 0 & 0.104\% & 2392648 & 15.639\% & 0 & N/A & 61440 & 66.260\% \\ \hline
    MP-SPECK  & 0.81\% & 0 & 0.051\% & 66369615 & 36.621\% & 0 & N/A & 0 & 0.108\% & 2392648 & 15.630\% & 0 & N/A & 61440 & 66.228\% \\ \hline
    \end{tabular}
}
\end{table*}
\begin{table}[!ht]
\vspace{-19pt}
\scalebox{0.88}{

\centering
\begin{tabular}{|l|l|l|l|l|}
\hline
\multirow{3}{*}{Hash} & \multicolumn{4}{c|}{Implementation Footprint}                                                 \\ \cline{2-5} 
                      & \multicolumn{2}{l|}{Clock per Byte} & \multirow{2}{*}{ROM usage} & \multirow{2}{*}{RAM usage} \\ \cline{2-3}
                      & Short.            & Long.           &                            &                            \\ \hline
BLAKE2s               & 3423              & 485             & 3606                       & 108                        \\ \hline
MD5                   & 1020              & 84              & 7328                       & 54                         \\ \hline
SHA-1                 & 4882              & 760             & 16518                      & 116                        \\ \hline
SHA-3                 & 79338             & 6217            & 1430                       & 410                        \\ \hline
BMW                   & 4801              & 311             & 17742                      & 138                        \\ \hline
DM-SPECK              & 9642              & 6020            & 1494                       & 70                         \\ \hline
MMO-SPECK             & 9520              & 6020            & 1542                       & 70                         \\ \hline
MP-SPECK              & 9642              & 6021            & 1542                       & 70                         \\ \hline
\end{tabular}
}\\
		\footnotesize{$^1$We assume that cryptographic Hash functions such as Blake2s, SHA-1, SHA-3 have been well designed and verified; here we only test MD5 as a baseline.}
\end{table}
 
\subsection{Non-cryptographic Hash Functions: Security Performance}
 Unlike the cryptographic hash functions, the hash functions built upon block ciphers might not necessarily pertain the desirable security properties of a hash function. In this context, we first assess their security related properties using statistical tests, where the test suite SMHasher\footnote{\url{https://github.com/aappleby/smhasher}} designed to evaluate security performance of non-cryptographic functions is adopted. The security performance is, in general, assessed through evaluating the distribution and the collision properties of the non-cryptographic hash functions. We briefly describe these tests below.
\vspace{1mm}


\noindent\textbf{Avalanche Tests: }The avalanche test evaluates the degree to which a hash value is affected by a single bit flipped in the message.
\vspace{1mm}

\noindent\textbf{Differential Tests: }For all possible small $n$-bit subsets of a $k$-bit key, generate 1,000 random key pairs that differ in only those $n$ bits and assess if any hashes to the same value; if so, the hash function is prone to far more collisions than expected.
\vspace{1mm}

\noindent\textbf{Keyset Tests:} 
A set of keys are generated, hashed and the resulting hash values are analyzed. The keyset tests generally assess: i) how evenly the hash values spread over the $2^n$-bit space; and ii) how often unrelated keys produce identical hash values. Overall, the keyset test results are measured by: i)~collisions; and ii) distribution.


{\bf Collision} (abbreviated as Col. in Table~\ref{tab:hash_result}), measures the degree to which more than one message shares the same hash value. Ideally, a hash function should be free of collisions, regardless of the format of the input message. In practice, the probability of finding a collision is given by $P_{col}$ in~\eqref{eqn:collisionRate}\footnote{https://preshing.com/20110504/hash-collision-probabilities/} where $k$ is the the number of  hashed values, and $N$ is the total number of possible hash values (for example, $N=2^{128}$ for a hash with a 128-bit binary output).

\begin{equation}
    P_{col} \approx 1-\exp{\bigg(\frac{-k(k-1)}{2N}\bigg)}
    \label{eqn:collisionRate}
\end{equation}


 Ideally, if one bit in the message vector is flipped, each bit in the hash output should have an equal probability (50\%) of flipping its value. In the {\bf distribution} measure (abbreviated as Dist. in Table~\ref{tab:hash_result}), 0\% implies that message bit information is perfectly distributed across all hash value bits, and 100\% implies that at least one hash value bit ether has a direct correlation with one specific message bit; or the hash value is not affected by any message bit. A high distribution value implies information leakage through the hash function.

\subsection{Results}
The tested results in terms of clock cycles per byte and memory footprint are summarized in Table \ref{tab:hash_result}. In terms of clock cycle performance, we can observe that standard cryptographic hash functions always outperformed the non-cryptographic counterparts constructed from the lightweight SPECK block cipher, while the MD5 exhibits the best performance among all tested hash functions. In terms of memory footprint, the non-cryptographic hash functions appears to have better performance than standard cryptographic counterparts.

However, SPECK non-cryptographic hash functions display notable collisions in the TwoBytes, Permutation and Zeros keyset tests. Meanwhile, non-cryptographic hashes tends to present large undesirable bias in TwoBytes and Zeros keyset tests. Although these results are not unexpected, we have found that the construction of a hash from the lightweight block ciphers we have evaluated have not delivered a lightweight hash function in a soft implementation.


Therefore, based on the set of evaluation measures we have employed, we can see that MD5 designed for software implementation provided the best \textit{overall} suitability (performance and security) for resource limited computational platforms. 

\section{Intermittent Execution Model}
\begin{figure}[!ht]
    \centering
    \includegraphics[width=0.85\linewidth]{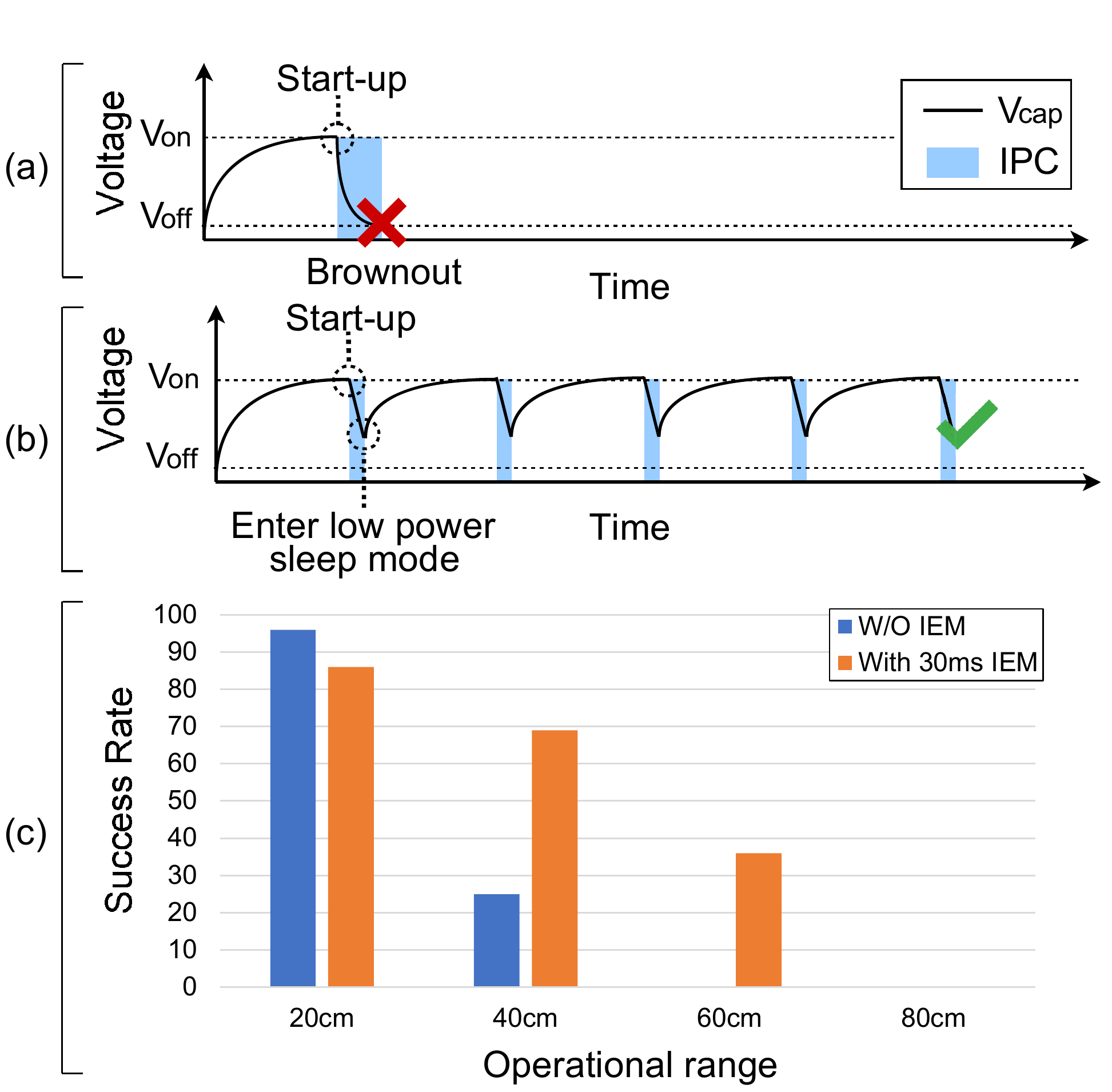}
    \caption{$V_{\rm cap}$ and intermittent power cycle (IPC) under (a) continuous operation mode and (b) intermittent operation mode. (c) The success rate of completing BLAKE2s-256 hash.}
    \label{fig:IEM}
\end{figure}

Commodity MCUs generally support multiple operational modes, including low-power (sleep) and power-down modes. The intermittent powering leads to limited clock cycles within an IPC, therefore, completion of e.g., hash, might not be possible within one IPC. In this context, a low-power sleep state~\cite{buettner2011dewdrop} can be used to intermit execution to allow the reservoir capacitor restore its energy. This intermittent execution model (IEM) can realize task execution in an energy limited setting~\cite{su2018secucode}. First, it stretches the time frame to complete the task and hence allows the energy harvester to collect more energy. Second, memory state is {\it preserved} during the low-power sleep mode. 
IEM is illustrated in Fig.~\ref{fig:IEM} (a-b) while the success rate for performing BLAKE2s hash over a 1,280~byte long message is shown in Fig.~\ref{fig:IEM}(c). We can see that the success rate at the 40~cm operating distance is nearly tripled, and the device could be used at a 60~cm range from an RFID reader antenna; which otherwise would be unfeasible without IEM.

\section{Conclusion and Future Work}
In this work we have presented benchmarks for \textit{eight software based hash function} implementations aimed at resource constrained devices. Our benchmarks are evaluated on a batteryless CRFID device. MD5 hash demonstrated the best overall performance, in particular, clock cycle overhead; thus, it is recommended\footnote{We remark here that in many applications involving resource limited devices, often, the security of the protocol impinges on the one-way property and we do not need the property of collision resistance.}.  In an energy constrained environment, we advocate employing IEM to overcome the problems posed by limited energy (clock cycles). Future work should consider a more detailed investigation of non-crytographic hash functions---including analyzing their security, optimizing code to reduce the implementation overhead, further investigating IEM settings along with their applicability and performance improvements. 
Moreover, we have not considered side-channel attacks, e.g., power- and timing-based, resilience when implementing codes and we leave this for future work.


\section*{Acknowledgement}
We acknowledge grant funding support from the Australian Research Council Discovery Program (DP140103448) and NJUST Research Start-Up Funding (AE89991/039).


\end{document}